\documentclass[11pt]{article}
\pdfoutput=1
\usepackage{graphicx}
\usepackage{cleveref}
\usepackage{amsmath}
\usepackage{amssymb}
\usepackage{slashed}
\usepackage{centernot}
\DeclareMathOperator{\Tr}{Tr}

\newcommand{\Fermi}{G_F}

\newcommand{\cc}[1]{\overline{#1}}
\newcommand{\weakf}{\frac{G_F}{\sqrt{2}}}
\newcommand{\lh}{(1-\gamma^5)}

\newcommand{\demi}{\frac{1}{2}}
\newcommand{\beq}{\begin{eqnarray}}
\newcommand{\eeq}{\end{eqnarray}}
\newcommand{\polp}{\mathbf{P}}
\newcommand{\lb}[1]{#1_{\beta}}

\begin{document}
\title {Rates and differential distributions in heavy neutral lepton
production and decays}
\author{Jean-Michel Levy \thanks{Laboratoire de Physique Nucl\'eaire et de Hautes Energies,
CNRS - IN2P3 - Sorbonne Universit\'e (Paris VI et Paris VII) and T2K collaboration.  \it Email: jmlevy@in2p3.fr}}
\maketitle
\vspace{1 cm}

\par
\section{Introduction}
The purpose of the following text is to show how one can calculate the production rate, polarization and differential decay distribution of a hypothetical "heavy neutrino" (or heavy neutral lepton, often abbreviated HNL in the following) with the couplings of ordinary fermions to standard model gauge bosons, up to a (very small) mixing matrix element. It is meant for experimental groups wishing to evaluate the sensitivity of their apparatus to HNL production and decay through simulation.\\
These HNL appear in various extensions of the standard model with massive neutrinos. The addition of right-handed Weyl spinors to the SM left-handed, massless, neutrinos and various hypotheses about
the mass matrices permit to envisage a large variety of physical states, among which some might have ordinary fermion masses, making them detectable through their decays into charged particles. In this paper, we will not dwelve much about those various mechanisms but will limit ourselves to the basic ingredients necessary to justify the possibility of these new fermions and of their couplings, in as much as these will be necessary for our purpose which is essentially to calculate their production and decay rates and their differential decay distributions required to evaluate the sensitivity of an experimental set-up.
  
It is important that masses have to be taken into account at every stage of (the simulation of) the production/decay process since, for example, the well known helicity suppression of $\nu_e$ production in two-body $0^-$ mesons decays no longer works when the neutrino is hypothetized to have a mass of a few MeV,
showing that rates can vary largely with mass. Also, polarization of the HNL must be taken into account because it bears on the differential distributions of its decay products and therefore on the acceptance of the experimental set-up to a given combination of mass and mode. Most results given here can probably be found in the litterature, see e.g. \cite{jml,JMB,GShap} but they are scattered among  many experimental or theoretical papers, which is why we think this one might have some usefulness. Moreover, some of these papers contain errors, see e.g. \cite{faux1}, which we came across while trying to help a young experimentalist colleague preparing his thesis \cite{mathieu}.\\
We will therefore give fairly complete derivations so as to allow anyone with a minimal litteracy in Dirac algebra to check our results. We apologize in advance to the many people who published such or such result for not quoting them. It is a task beyond our capacity and anyway, quite useless in a work of this kind. \\
The present paper is devoted to "heavy neutrino" production by decay of charged $0^-$ mesons
and is restricted to 2-body ($0^-$ charged meson plus charged lepton) and the simplest 3-body not involving neutral currents (non self-conjugate charged lepton pair plus light neutrino). A second paper will be devoted to the production/decay of HNL through neutral currents which needs some investigation of the later
in the neutrino sector of the extensions of the SM envisionned.

\section{Generalities - effective lagrangian}
Neutrino states related to charged leptons through weak charged currents are
thought to be linear combinations of mass eigenstates. The mechanism giving rise to
these combinations is not known, but given the successes of "standard" physics,
we assume that the interaction lagrangian is that of the Standard Model, namely:
$${\cal L}_{int} = e A^{\alpha}J^{em}_{\alpha} + \frac{g}{cos\theta_w}
Z^{\alpha}J^{neut}_{\alpha} + \frac{g}{\sqrt{2}}(W^{\alpha\dag
}J_{\alpha}^{ch}+W^{\alpha
}J_{\alpha}^{ch\dag}) $$ 
\footnote{Einstein's summation convention is used thourough for space-time indices}
where: \\
$$J^{ch}_{\alpha} = \ \sum_{\beta = e,\mu,\tau} \
\overline{\nu}_{\beta}\gamma_{\alpha}{\cal P}_L l_{\beta} + quark \ currents$$ 
$$J^{neut}_{\alpha} = \ \sum_{f} \ \overline{f}\gamma_{\alpha}({\cal P}_LT^3_w
-sin^2\theta_w Q)f $$
$$J^{em}_{\alpha} = \ \sum_{f}\ \cc{f}\gamma_{\alpha}Qf$$
\begin{itemize}
\item $f$ is any elementary fermion field, $\nu_{\beta}$ and $l_{\beta}$ stand
for the neutrino and charged lepton fields of "flavour" $\beta$ ($= e, \mu, \tau$).
\item $T^3_w$ and $Q$ are the third weak isospin component and electric charge operators.
\item ${\cal P}_L = \demi \lh$ is the left-handed projector.
\item $g = \frac{e}{sin\theta_w}$
\end{itemize}
\par
$\nu_{\beta}$'s are assumed to be linear superpositions of
fields corresponding to definite mass quanta which can be either Dirac or Majorana.
The notation will be as follows:
\beq \nu_{\beta} = \sum_h \ U_{\beta h} N_h \label{super} \eeq
where $N_h$ represents the field of a neutrino of mass $m_h$. Greek indices will be used for leptonic "flavours" and latin indices for definite mass fields. Results of neutrino oscillation experiments demonstrate that the superposition on the RHS of equation (\ref{super}) comprises at least three fields with different masses and -baring the case of extreme degeneracy- these masses must be much smaller than those of charged fermions.  However, nothing forbids the existence of higher mass states, provided the relevant $U$ matrix element is sufficiently small for having kept the corresponding state 'incognito' in the experiments so far performed. In the following, we will assume that there is at least an extra "heavy neutrino" (Heavy neutral lepton or HNL henceforth). $U$ is therefore an extension of the usual PMNS mixing matrix (the unitarity of which has not been experimentally verified). We will evidently assume that $U$
is square and unitary, so that relation (\ref{super}) is simply inverted through:
\beq N_h = \sum_{\beta} \ U^*_{\beta h} \nu_{\beta} \eeq
\par
The processes of interest are at low energies and will always involve virtual 
$W$ and $Z$'s. Therefore, they will be at least second order in ${\cal L}_{int}$. Neglecting $q^2$ w.r.t. $m^2$ in the bosons propagators written in momentum space, one finds the effective lagrangian:
\begin{eqnarray}
{\cal L}_{eff} = \frac{4
G_F}{\sqrt{2}}(J^{neut,\alpha}J_{\alpha}^{neut}+J^{ch,\alpha} J_{\alpha}^{ch
\dag})  \label{effl}
\end{eqnarray}
with $\frac{G_F}{\sqrt{2}} = \frac{g^2}{8M_W^2}$  in tree approximation.
\par
In the following, we shall note $j$ for a leptonic current and $J$
for a hadronic current.
\section{Production through 2-body $0^-$ charged mesons decay}
The relevant part of the effective lagrangian is here:
$$\frac{4 G_F}{\sqrt{2}}(j^{ch,\alpha}J^{ch \dag}_{\alpha}+h.c.)$$
From now on, we assume a $M^+$ (momentum $P$, mass $M$) decaying to HNL $N_h$  (momentum $p_h$, mass $m_h$), and antilepton $\beta^+$ of "flavour" $\beta$ (momentum $\lb{p}$, mass $\lb{m}$). $M^+$ being spinless, the only vector available to parametrize the hadronic current matrix element is its 4-momentum $P^{\alpha}$.\footnote{Further notice that only the axial part of the hadronic current can have a non-zero matrix element between a pseudoscalar state and the hadronic vacuum.} 
Introducing $M^+$ 'decay constant' $f_{M}$ and using Lorentz invariance one makes the usual {\it ansatz} for the current matrix element:Rates and differential distributions in heavy neutral lepton
production and decays
$$<O|A^{ch\;\alpha\;\dag}(x)|M^+> = if_{M} V_{..}e^{-iP \cdot x}P^{\alpha}$$ 
where $V_{..}$ is the relevant CKM matrix element for $M^+ \rightarrow {W^+}^*$ \footnote{${W^+}^*$ is an off-shell $W^+$}.
\par
The leptonic current matrix element for the $M^+ \rightarrow N_h  \beta^+$ decay is: 
$$<\beta^+\;N_h|\sum_{k,\delta} U^*_{\delta k}\overline{N_k}\gamma_{\alpha}{\cal P}_Ll_{\delta} (x)|O>  =
U^*_{\beta h}\overline{u}(N_h)\gamma_\alpha{\cal P}_Lv(\beta)
e^{i(\lb{p}+p_h)\cdot x}$$
so that the transition matrix element will be:
$$-i\sqrt{2}G_Ff_{M}U^*_{\beta h}V_{..}\overline{u}(N_h)\slashed{P}(1-\gamma^5)v(\beta)
\;\;\footnote{${\slashed P}$ stands for $P^{\alpha}\gamma_{\alpha}$ (Feynman
's notation.)}$$
\par
This result is obviously independant of the Dirac or Majorana nature of the $N_h$
field.
\par
As said in the introduction, we give here a detailed derivation, only assuming that the reader knows how to calculate traces of products of Dirac algebra matrices. Our way of calculating the HNL polarization vector and using it in the second decay is inspired by \cite{Landau}
\begin{enumerate}
\item using $P= p_h^{}+\lb{p}^{}$ and the Dirac equations:\\

 $\bar{u}\slashed{p}_h = m_h^{}\bar{u}\;\;$ and $\;\;\lb{\slashed{p}}v = -\lb{m}^{} v$\\

simplify the matrix element to $-i\kappa\bar{u}(a-\gamma^5b)v$ \\
where $\kappa =\sqrt{2}G_Ff_{M}U^*_{\beta h}V_{..}$, $a = m_h^{}-\lb{m}^{} \;,\; b = m_h^{}+\lb{m}^{}$
\item multiply the m.e. by its complex conjugate:\\
 
$\kappa^2\bar{u}(a-\gamma^5b)v\bar{v}(a+\gamma^5b)u$\\

$ = \kappa^2\Tr(u\bar{u}(a-\gamma^5b)v\bar{v}(a+\gamma^5b)$

\item sum over antilepton polarizations, which amounts to the replacement: \\

$v\bar{v} \rightarrow (\lb{\slashed{p}}-\lb{m}^{})$.\\

\item In order to calculate the HNL polarization, keep its full density matrix for both momentum and spin: \\

$u\bar{u}\rightarrow(\slashed{p}_h+m_h^{})\demi(1+\gamma^5\slashed{s})$ \\

where $s$ is the HNL polarization 4-vector which reduces, in the rest frame, to $(0, \mathbf{P})$ with $\mathbf{P}$ the usual polarization 3-vector for spin 1/2, i.e. twice the spin expectation value. 

\item the squared m.e. thus becomes: \\

$\kappa^2\Tr{(\slashed{p}_h+m_h^{})\demi(1+\gamma^5\slashed{s})(a-\gamma^5b)(\lb{\slashed{p}}-\lb{m}^{})(a+\gamma^5b)}$\\

 
\item Calculate the trace. Using again 4-momentum conservation, this yields:\\
\begin{eqnarray} 1/4\Tr = M^2(\lb{m}^2+m_h^2)-(\lb{m}^2-m_h^2)^2+2m_h(m_h^2-\lb{m}^2)s\cdot p_{l} \label{me2}\end{eqnarray}
\end{enumerate}

\begin{figure}[h] 
\begin{center} 
\includegraphics[width=\linewidth]{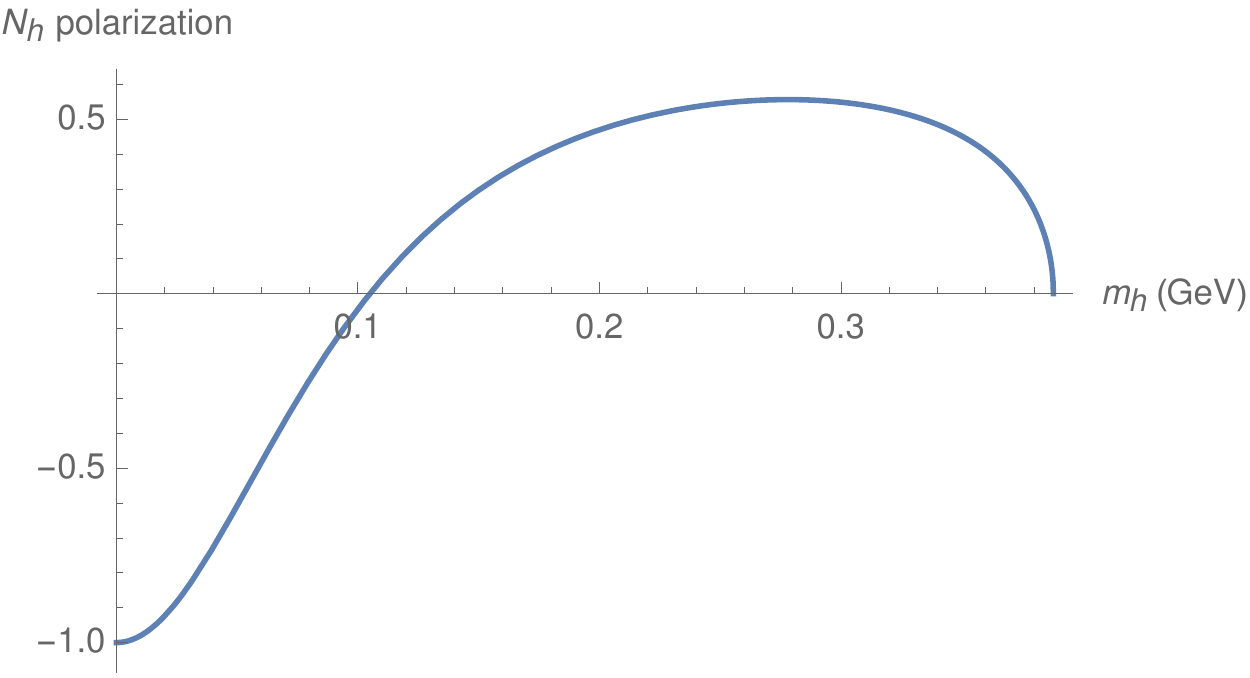} 
\caption{\small Polarization of HNL produced in $K^+ \rightarrow {\rm HNL} + \mu^+$ as a function of HNL mass. It is seen that when the latter coincides with the muon mass, the polarization vanishes} 
\label{fig: cnufepi} 
\end{center} 
\end{figure}

\begin{itemize}
\item To calculate the rate, sum over HNL spin states by replacing $s \rightarrow 0$ and multiplying
by 2.\\

Adding normalization and phase-space factors, one gets the width:

$$\Gamma(M^+\rightarrow \beta^+ N_h) = \frac{G_F^2f^2_{M}|V_{..}|^2|U_{\beta h}|^2}
{8\pi M}\left (m_h^2+\lb{m}^2-\frac{(\lb{m}^2-m_h^2)^2}{M^2}\right )\lambda^{1/2}(M^2,m_h^2,\lb{m}^2)$$

where $M, \lb{m}, m_h$ are the masses of $M^+, \beta$ and $N_h$ respectively and $\lambda$ 
is the usual kinematical function $$\lambda(x,y,z)=x^2+y^2+z^2-2(xy+yz+zx)$$
By letting $m_h \rightarrow 0$ (and forgetting about $U$) one retrieves the ordinary well known
formula for decay into an antilepton and a standard model massless neutrino.
\begin{eqnarray}
\Gamma =
\frac{G_F^2f^2_{M}|V_{..}|^2}{8\pi }M\lb{m}^2(1-\frac{\lb{m}^2}{M^2})^2
\end{eqnarray} 
which, being proportionnal to $\lb{m}^2$, explains the tiny ratio $\Gamma(e^+\nu)/\Gamma(\mu^+ \nu)$,  
due to helicity conservation by V and A  vertices in the ultra-relativistic limit. For the domain envisioned here ($\mu \geq \rm{\;a\; few\;} MeV$) the suppression no longuer works and both modes acquire the same order of magnitude modulo the coefficients $|U_{\beta h}|$ 
\item To find the HNL polarization:\\

The squared m.e.(cf. \ref{me2}) 
is proportionnal to the probability of finding 4-polarization $s$ and must therefore be equal to $\Tr(\rho\rho_f)$ (with $\rho_f$ the true HNL polarization matrix) up to a factor.
\\ In the rest frame, $\rho$ reduces to $\mathbf{1+\sigma \cdot \polp}$ with $\mathbf{\sigma}$ the Pauli matrices and $\polp$ the polarization 3-vector, so that the expression obtained is proportionnal to $\Tr{\mathbf{(1+\sigma\cdot\polp)(1+\sigma\cdot\polp^f)}}$ or to $\mathbf{1+\polp\cdot\polp}^f$

\par
By expliciting the proportionality of this last expression with (\ref{me2}) written in the HNL rest frame, we find the following for the HNL polarization vector to be used when simulating its decay:
\begin{eqnarray} \label{polar}
\mathbf{P} = \frac{(\lb{m}^2-m_h^2)\lambda^{1/2}(M^2,m_h^2,\lb{m}^2)}{M^2(\lb{m}^2+m_h^2) - (\lb{m}^2-m_h^2)^2}\mathbf{\hat{n}} = \mathbb{P}\mathbf{\hat{n}}
\end{eqnarray}
where $\mathbf{\hat{n}}$ is a unit vector in the direction of the parent meson or of the decay lepton in the $N_h$ rest frame and the second equality defines $\mathbb{P}$. Although the formula obtained by the authors of \cite{faux1} is not given in their paper, it is readily seen graphically (compare with fig. \ref{fig: cnufepi})that it must coincide with our above result for the case where the initial particle is a charged kaon decaying into muon and HNL. In particular, it is seen from the graph (fig. \ref{fig: cnufepi}) or formula (\ref{polar}), that if the HNL mass $m_h$ coincides with the muon mass, its polarization vector is zero.
The graph or formula (\ref{polar}) also show that when $m_h \rightarrow 0$, the coefficient of $\mathbf{\hat{n}}$ $ \rightarrow 1$  that is, the massless neutrino will be pure $-1$ helicity.   
\end{itemize}
\pagebreak
\section{Two-body HNL decay into $0^-$ meson and lepton}
These are crossed channels of those envisionned above for production. Amplitudes 
are trivial to write; in the 'charged' case, one finds e.g., for a decay into $\pi^+ , \beta^-$:\\
(Here $k$, $q$ and $p$ are the 4-momenta of $N_h$( mass $m_h$), $\beta^-$ (mass $\lb{m},\;\rm{flavour\; }\beta)$ and $\pi^+$ (mass $m_{\pi}$) so that $ k = p+q$) 

$A_{N_h\rightarrow \pi^+ l^-} = -i\frac{G_F}{\sqrt{2}} f_{\pi} V_{u,d}U_{h,\beta}^{} \bar{u}(l)\slashed{p}(1-\gamma^5)u(N_h)$ \\

 Using Dirac equation, one gets:\\ 
$$A_{N_h\rightarrow \pi^+ l^-} = -i\frac{G_F}{\sqrt{2}} f_{\pi}V_{u,d}U_{h,\beta}^{} \bar{u}(l)(a+b\gamma^5)u(N_h)$$ 

with $a = m_h -\lb{m}^{}$ and $b = m_h + \lb{m}^{}$\\

Squaring, summing over $l$ polarizations and introducing the HNL polarization matrix with polarization 4-vector $s$ one gets:\\
$$\overline{|A|^2} = \frac{G_F^2}{2}f_{\pi}^2|U_{h,\beta}^{}|^2|V_{u,d}|^2 \Tr{(\slashed{q}+\lb{m})(a+b\gamma^5)(\slashed{k}+m_h)\demi(1+\gamma^5\slashed{s})(a-b\gamma^5)}$$

Calculating the trace, one gets \\
$$1/4 \Tr{} = (a^2+b^2)q\cdot k + 2m_h\,ab\,q\cdot s +\lb{m}^{}m_h(a^2-b^2)$$
$$=(m_h^2-\lb{m}^2)^2-m_{\pi}^2(m_h^2+\lb{m}^2)+2m_h(m_h^2-\lb{m}^2)q\cdot s\;\footnote{We have used $2q\cdot k = m_h^2+\lb{m}^2-m_{\pi}^2$}$$
and $q\cdot s$ reduces to $-\mathbf{q}\cdot\polp$ 
in the $N_h$ rest frame, therefore:
\begin{eqnarray}\overline{|A|^2} = G_F^2f_{\pi}^2 |U_{h,\beta}|^2|V_{u,d}|^2 [{(m_h^2-\lb{m}^2)^2-m_{\pi}^2(m_h^2+\lb{m}^2)-2m_h(m_h^2-\lb{m}^2)\mathbf{q}\cdot\polp}]  \label{mesquared} 
\end{eqnarray}

With phase space (integrated over the angles)  equal to $\frac{|\mathbf{q}|}{4\pi m_h}$ or
$\frac{\lambda^{1/2}(m_h^2,m_l^2,m_{\pi}^2)}{8\pi m_h^2}$ we find for the rate: 
$$\Gamma(N_h \rightarrow \beta^-\pi^+) = \frac{G_F^2f_{\pi}^2}{16\pi m_h^3}
\{(m_h^2-\lb{m}^2)^2-m_{\pi}^2(m_h^2+\lb{m}^2)\}\lambda^{1/2}(m_h^2,\lb{m}^2,m_{\pi}^2)|U_{h,\beta}|^2|V_{u,d}|^2$$
If $\beta=e$ , $\lb{m}$ can be neglected and this becomes:$$\frac{G_F^2f_{\pi}^2}{16\pi}m_h^3\left(1-\frac{m^2_{\pi}}{m_h^2}\right)^2|U_{h,e}|^2|V_{u,d}|^2$$

The angular distribution is anisotropic due to polarization (cf. (\ref{polar})) as evidenced by (\ref{mesquared}) 

Normalizing formula (\ref{mesquared}) (so that the integral over $\cos{(\mathbf{\hat{n}}, \mathbf{\hat{q}})}$ equals 1 ) we get:
$$\frac{dN}{d\cos{\theta}} = 1/2 - 1/2\frac{m_h^2-\lb{m}^2}{(m_h^2-\lb{m}^2)^2-m_{\pi}^2(m_h^2+\lb{m}^2)}\lambda^{1/2}(m_h^2,\lb{m}^2,m_{\pi}^2) \mathbb{P}\cos{\theta}$$

$\theta = (\mathbf{\hat{n}},\mathbf{\hat{q}})$ is the angle between the recoil lepton direction ($\mathbf{\hat{n}}$) in the parent's decay $M^+ \rightarrow \rm{HNL} + \beta^+$ and the secondary lepton direction ($\mathbf{\hat{q}}$) due to HNL decay, seen in the HNL rest frame. $\mathbb{P}$ has been defined  in (\ref{polar}). \\
It is clear that, contrary to formula (16) of ref.\cite{faux1} the HNL decay is isotropic
when its mass equals that of the lepton recoiling against it in the parent's decay. This formula is incoherent on different other grounds, making, for example, no distinction between the c.o.m. momenta in the HNL-generating meson two-body decay and the HNL two-body decay itself.

\subsection{A pedagogical remark}
It is interesting to note that the heavy neutrino which is in general only partially polarized is NOT a quantum mechanical linear superposition of helicity 1 and helicity -1 states contrary to what is stated in many places (see e.g. \cite{Petcov}). Since its polarization vector modulus is not one, there is no direction in which a spin measurement will yield 1/2 with certainty and this system, which is in a mixed state, cannot be represented by a wave function. Although the spin 0 initial meson can be thought of as being in a pure state, the HNL, being but a subsystem of the -evolved- initial state can only be represented by a density matrix (see e.g.\cite{Landau2})

\section{Decay into a light neutrino and a non charge-conjugate lepton pair ($\beta^-\;\beta'^+\;\nu)$}
\subsection{Decay matrix element}
\includegraphics[width=\linewidth]{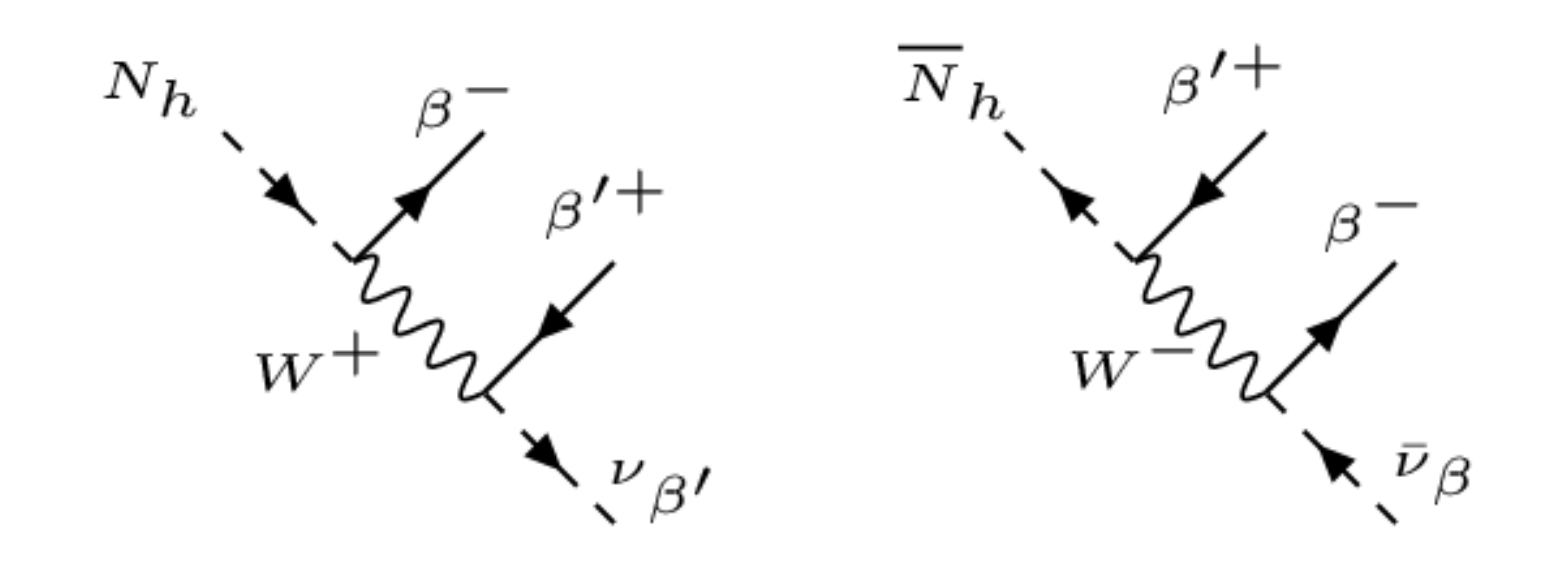} 
For a Dirac HNL and in tree approximation, the decay is described through the left-hand-side Feynman diagram here above. \footnote{For clarity, we have 'uncollapsed' the $W$ propagators in the diagrams}

Only the $\beta$ component of $N_h$ contributes to the first vertex so that a factor of $U_{\beta,h}^*$ enters the matrix element at
this level. For the final state with a charged lepton of flavour $\beta'$,
the final neutrino also has flavour $\beta'$. Obviously, for kinematical calculations not involving oscillations as we are dealing with here, these neutrinos can be considered to be massless and no mixing matrix elements need be introduced at this level.

On the other hand, for a Dirac anti-neutrino, it is the right hand side diagram that must be calculated,
and as above, the final neutrino mass is neglected and no $U$ factor is needed for the final state. 

If neutrinos are Majorana particules, both diagrams must be considered, but due to the different
flavours of the final neutrinos, no interference can take place.

\par
The relevant part of the effective lagrangian is now : $${\cal L'} =
\frac{4G_F}{\sqrt{2}}\sum_{k k' \alpha \alpha'} U_{\alpha k}U^*_{\alpha' k'}
\cc{l}_{\alpha}\gamma^{\mu}{\cal P}_LN_k\cc{N}_{k'} \gamma_{\mu}{\cal P}_L l_{\alpha'} $$
If the $N_h$ are Dirac fields, the transition amplitude is simply:
$$4\weakf U_{\beta h} \cc{u}_{\beta}\gamma^{\mu}{\cal P}_L u_h \cc{u}_{\beta'}^{\nu}\gamma_{\mu}{\cal P}_L v_{\beta'}$$
here $u$'s and $v$'s are Dirac spinors and ${\cal P}_L$ is the projector on their left-handed part.\\
In conformity with the remark made about the final state neutral lepton, the $U^*_{\beta' k'}\cc{N}_{k'}$ sum has been replaced by the sole $\cc{u}_{\beta'}^{\nu}$ standing for a SM neutrino of flavour $\beta'$ produced together with the $\beta'^+$ charged anti-lepton.

For the Majorana case, there will be the extra piece:
$$4\weakf U_{\beta' h}^* \cc{u}_{\beta}\gamma^{\mu}
{\cal P}_Lv_{\beta}^{\nu} \cc{v}_h\gamma_{\mu}{\cal P}_Lv_{\beta'}$$
with an analogous remark for the absence of final neutrino mixing matrix element and for the spinor 
$v_{\beta}^{\nu}$ which stands now for the light neutrino of flavour $\beta$ produced together with the $\beta$-flavoured charged lepton. \\

We now Fierz-transform these amplitudes (see ref. \cite{Landau}, \textsection\space 28 ) so as to render both first factors equal, getting: 
$$-4\weakf U_{\beta h} \cc{u}_{\beta}\gamma^{\mu}{\cal P}_L v_{\beta'}\cc{u}_{\beta'}^{\nu}\gamma_{\mu}{\cal P}_L u_h  $$
and
$$-4\weakf U_{\beta' h}^* \cc{u}_{\beta}\gamma^{\mu}
{\cal P}_Lv_{\beta'}\cc{v}_h\gamma_{\mu}{\cal P}_Lv_{\beta}^{\nu} $$
and we use the relation $$\cc{v}_h\gamma^{\mu}(1-\gamma^5) v_l = \cc{u}_l\gamma^{\mu}(1+\gamma^5)u_h$$
in order to have (almost) the same spinors sandwiching the second factors: indeed, $l$ stands for $\beta$ or $\beta'$ which correspond to orthogonal states, but the spinors are identical four-components functions of the momentum variable.\\ 
\subsection{Differential distribution}
This being done, the two amplitudes can be added, yielding:
$$ -2\weakf \cc{u}_{\beta}\gamma^{\mu}{\cal P}_Lv_{\beta'} \cc{u}_l\gamma_
{\mu}((U_{\beta h}+ U^*_{\beta' h})-(U_{\beta h}-U^*_{\beta' h})\gamma^5)u_h$$ 

where it is understood here that since $\beta \neq \beta'$ interference terms (containing products like $U_{\beta h}U^*_{\beta' h}$) should be cancelled in the end, rendering the question of the relative sign of the two amplitudes irrelevant. Moreover, since the final neutrino helicities are opposite, one does not expect such terms to occur. \\

To simplify let $U_{\beta h}+ U^*_{\beta' h} = \alpha\;\;  ,\;\; U_{\beta h}-U^*_{\beta' h} = \beta$. 
Squaring the above expression, we get:
$$2\Fermi^2 \cc{u}_{\beta}\gamma^{\mu}{\cal P}_Lv_{\beta'}\cc{v}_{\beta'}\gamma^{\nu}{\cal P}_Lu_{\beta}\cc{u}_l\gamma_{\mu}(\alpha-\beta\gamma^5)u_h\cc{u}_h\gamma_{\nu}(\alpha^*-\beta^*\gamma^5)u_l$$ which we rewrite, summing over final polarizations and introducing the HNL density matrix:
$$2\Fermi^2 \Tr{((\slashed{p}_-+m_{\beta})\gamma^{\mu}{\cal P}_L(\slashed{p}_+-m_{\beta'})\gamma^{\nu}{\cal P}_L)}\Tr{(\slashed{q}\gamma_{\mu}(\alpha-\beta\gamma^5)(\slashed{k}+\mu_h)\demi(1+\gamma^5\slashed{s})\gamma_{\nu}(\alpha^*-\beta^*\gamma^5))}$$
where $k,p_-,p_+,q,s$ are the 4-momenta of $N_h,\beta^-,\beta'^+,\nu_l$ and the $N_h$
 4-polarization. $m_h$, $m_{\beta}$ and $m_{\beta'}$ are the masses of $N_h$ and of the two charged leptons.  Taking the traces and contracting the Lorentz indices then yields:
\begin{eqnarray}
64G^2_F(|U_{\beta h}|^2q\cdot p_-(k-m_hs)\cdot p_+ + |U_{\beta' h}|^2 q\cdot 
p_+(k+m_hs)\cdot p_-) \label{4dim}
\end{eqnarray}

As expected, no spurious interference terms need to be cancelled explicitly.\\

The last expression can easily be transformed to:
\begin{eqnarray}
64G^2_Fm_h^2(|U_{\beta h}|^2(E^*_+-E_+)(E_++ \mathbf{P}\cdot {\bf p}_+)+|U_{\beta' h}|^2(E^*_--E_-)(E_--\mathbf{P}\cdot {\bf p}_-))    \label{differential} 
\end{eqnarray}here: $E^*_\mp = (m_h^2 \pm m^2_{\beta} \mp m^2_{\beta'})/(2m_h)$ and $E_\mp$ , ${\bf p}_\mp $ are $\beta^-$ and $\beta'^+$ energies and 3-momenta in the decaying HNL rest-frame and 
$\mathbf{P}$ is its 3-polarization vector as calculated in Part I.\\

The three-body final state phase space depends on five variables only which can be taken, in the HNL center of mass frame, as $E_+$, $E_-$ and three angles defining the final state orientation. By energy-momentum conservation, the three final momenta are coplanar in this frame and the angle $\theta_{+-}$ between ${\bf p}_+$ and ${\bf p}_-$ is fixed once $E_+$ and $E_-$ are given \footnote{One finds: $2p_+p_-\cos{\theta_{+-}} = m_h^2+m_+^2+m_-^2 -2m_h(E_++E_-)+2E_+E_-$}. One can then choose the polar angles of ${\bf p}_+$ with respect to the HNL parent direction $\mathbf{\hat{n}}$, which is itself parallel to $\mathbf{P}$ (see (\ref{polar})),  call them $\theta_+$ and $\phi_+$ and the angle of the decay plane around ${\bf p_+}$ , say $\Phi$, to completely define the final state. In order to use formula (\ref{differential}), one only needs the cosine of the angle of ${\bf p}_-$ and $\mathbf{\hat{n}}$ which is found to be
\begin{eqnarray}
\cos{\theta_-} = \cos{\theta_+}\cos{\theta_{+-}}+\sin{\theta_+}\sin{\theta_{+-}}\cos{\Phi} \label{strig}
\end{eqnarray}
by a standard spherical trigonometry formula (see e.g. \cite{sphetrig}) in the spherical triangle defined by $({\mathbf{\hat{n}}, {\bf p_+}, {\bf p_-}})$ \\
     
Note that (\ref{differential}) is valid for a Majorana neutrino. For a Dirac neutrino going to $\beta^-,\beta^{'+}$,  the second term must be dropped and conversely, the first term must be dropped for a Dirac anti-neutrino decaying into the same \underline{charged} channel.

This result is again very different from those of \cite{faux1}, which nowhere gives the full differential decay distribution necessary for a proper simulation.\\
Observe however, that in order to use formula (\ref{differential}) to estimate the acceptance of the apparatus to the channel studied, some estimate of the ratio  $|U_{\beta h}|^2/|U_{\beta' h}|^2$ will have to be used.

\subsection{Decay width}
Practically, since $\beta \neq \beta'$,
one has e.g. $m_{\beta} = m_e \ll m_{\mu} = m_{\beta'}$ so that $m_{\beta}$ will 
be neglected.
\par
With this approximation, the width can be analytically integrated with the results:
\begin{eqnarray}
\Gamma = \frac{G_F^2m^5_h}{192\pi^3}\left \{|U_{\beta h}|^2+|U_{\beta' h}|^2\right \}f(r) \label{Majo}
\end{eqnarray} 
 Here: $r = (m_{\beta'}/m_h)^2$ and $f(r)=(1-8r+r^2)(1-r^2)-12r^2Log(r)$
\par
(\ref{Majo}) is valid for Majorana's neutrinos. The remarks already made above concerning the Dirac case apply. \\

For neutrinos produced by pions or kaons decays, the only kinematically 
allowed case is $\mu^{\mp} e^{\pm} \nu_l$.  
 Therefore, the channel $N_h \rightarrow \mu^- e^+ \nu_l$ yields a measure 
of $|U_{\mu h}|^2$ for Dirac neutrinos and $N_h \rightarrow \mu^+ e^- \nu_l$
 measures $|U_{e h}|^2$. For Dirac  anti-neutrinos, the channels are permuted. Lastly, for Majorana neutrinos, the sum $|U_{\mu h}|^2 + |U_{e h}|^2$ is measured by either 
channel.

\end{document}